
\input amstex
\magnification=1200
\TagsOnRight
\def\wh{\widehat}

\def\D{\Cal{D}}

\def\ov{\overline}

\def\noi{\noindent}

\overfullrule=0pt

\def\mapab#1{\Big\downarrow}

\mathsurround=1pt
\tolerance=10000
\pretolerance=10000
\nopagenumbers
\headline={\ifnum\pageno=1\hfil\else\hss\tenrm -- \folio\ --\hss\fi}
\hsize=16 true cm
\vsize=23 true cm
\line{Preprint {\bf SB/F/95-231}}
\hrule
\vglue 1.5cm
\vskip 2.cm

\centerline{\bf THE GAUGE INVARIANT LAGRANGIAN FOR }
\centerline{\bf SEIBERG-WITTEN TOPOLOGICAL MONOPOLES }

\vskip 1.5cm
\centerline{  R. Gianvittorio, I.Martin and A. Restuccia}
\vskip .5cm
\centerline{\it Departamento de F\'{\i}sica, Universidad Sim\'on Bol\'{\i}var}
\centerline{\it Apartado postal 89000, Caracas 1080-A, Venezuela.}
\centerline{\it and}
\centerline{\it International Center for Theoretical Physics, Trieste, Italy}

\vskip 2.0cm
\centerline{\bf Abstract}
\vskip .5cm
{\narrower{\flushpar  A topological gauge invariant lagrangian for
Seiberg-Witten monopole equations is constructed. The action is invariant
under a huge class of gauge transformations which after BRST fixing leads
to the BRST invariant action associated to Seiberg-Witten monopole
topological theory. The supersymmetric transformation of the fields
involved in the construction is obtained from the nilpotent BRST algebra.
\par}}

\vskip 3cm
\hrule
\bigskip
\centerline {\it e-mail: ritagian{\@}usb.ve , isbeliam{\@}usb.ve ,
arestu{\@}usb.ve }

\newpage

The recent Seiberg-Witten monopole equations [1] have risen high
expectations in both mathematics and theoretical high energy physics. On
one hand, the Seiberg-Witten theory provides new means of classifying
differentiable 4-manifolds, even when the complete equivalence between
the Donaldson polynomials and the Seiberg-Witten invariants  has not yet
been shown. On the other, it shades a new light on the duality problem of
Quantum Field Theory.

The $SU(2)$ topological quantum field theory of Witten [2] can be obtained
from a twisted version of N=2 supersymmetric Yang-Mills theory which
arises directly by BRST gauge fixing of a Lagrangian involving only the
curvature of the $SU(2)$ connection and an auxiliary two form [3]. The
supersymmetric transformations may be obtained directly from the BRST
algebra. As a consequence of the twisting there is no special requirement
over the spin structure on the general differentiable 4-manifold and the
quantum theory may be formulated starting from a general orientable riemmanian
4-manifold. However, the construction of a gauge invariant action for
Seiberg-Witten monopole equations requires from the beginning the existence
of a spin structure over the 4-manifold, luckily this existence is assured
for any orientable riemmanian manifold in four dimensions . In the case when
the second Stiefel-Whitney class of the 4-manifold is zero, i.e.
$\omega_{2}=0$, the $SO(n)$ structure group of the tangent bundle can
always be lifted to $Spin(n)$ and, hence, to define the corresponding spin
structure. In other cases when $\omega_{2}$ is reducible modulo two of an
integral cohomology class $c_1 \in H^2(X,Z)$, it is always possible to lift
$SO(n)$ to $Spin_{c}(n)= Spin(n) \times_{Z_{2}} U(1)$ and to define a
$Spin_{c}$ structure. As said before, over any orientable 4-manifold
 a $Spin_{c}$ structure can always be constructed as
$\omega_{2}$ is
always reducible modulo two of the integer Chern class [4]. This property
is not valid in general for manifolds of dimension $d > 4$ but holds
perfectly for orientable 4-manifolds. It is this unique property which
allows the Seiberg-Witten construction over a general riemmanian
4-manifold. The riemmanian requirement as we shall see arises only at the
level of fixing the gauge of our gauge invariant action.

In this article we introduce a topological action (1), invariant under a
huge class of local symmetries which after BRST fixing reduces to the
Seiberg-Witten theory. One of the main consequences of the existence of
this action would be the possibility of relating the $SU(2)$ topological
quantum field theory [2] directly to the Seiberg-Witten theory [1]. In
fact, the action (1) could be obtained by a partial gauge fixing that
breaks the $SU(2)$ invariance to a $U(1)$ in the action already obtained in
[3] for the $SU(2)$ topological theory. This would allow to compare
directly the correlation functions of both topological theories by using
the BFV theorem. This procedure seems interesting since does not use
explicitly the duality relation between both theories found in [1] nor it
does use any supersymmetric argument , the gauge actions (1) and the one
found in [3] are not supersymmetric.

The action over a general differentiable 4-manifold X we propose is given by
$$
S={1\over{4}}\int_X
(F_{\mu\nu}+B_{\mu\nu}+
{i\over{2}}\ov{M} \Gamma _{\mu\nu}
M)(F_{\rho\sigma}+B_{\rho\sigma}+
{i\over{2}}\ov{M} \Gamma _{\rho\sigma} M)dx^{\mu} \wedge
dx^{\nu} \wedge dx^{\rho} \wedge dx^{\sigma} , \tag1
$$
 The field  $F_{\mu\nu}$ is the curvature
associated to the
$U(1)$ connection $A_{\mu}$ over a complex line bundle $L$,
$B_{\mu\nu}dx^{\mu} \wedge dx^{\nu}$ is an
independent auxiliarly 2-form. $M$ and its complex conjugate $\ov{M}$ are
sections of $S^+\otimes L$ and $S^-\otimes L^{-1}$ respectively, where
$L^{-1}$ is the complex conjugate bundle of $L$ and $S^+$ is one of the
irreducible parts of the spinor bundle $S$. For any
even manifold with a $Spin_c$ structure there is always a unique spinor
bundle $S$ associated to a representation of $Spin_c$ that splits into a
direct sum $S(X)= S^+(X)\oplus S^-(X)$. The Clifford
matrices $\Gamma_{\mu}$ satisfy $\{\Gamma_{\mu}, \Gamma_{\nu}\}=2
g_{\mu\nu}$.
The action (1) is independent
of the metric, consequently its associated partition function is also
independent of it and the observables of the quantum theory are going to
be topological invariants.

The  local symmetries of this action are given by the following
infinitesimal transformations:
$$
\matrix
\delta_{\lambda} A_\mu=\D_{\mu}\Lambda,&\delta_{\lambda}B=0,&
\delta_{\lambda}M=0;
\endmatrix
\tag2
$$

$$
\matrix
\delta_{\epsilon}A_{\mu} = \epsilon_{\mu},&\delta_{\epsilon}M=0,
&\delta_{\epsilon}B_{\mu\nu} = -\D_{[\mu}\epsilon_{\nu{}]};
\endmatrix
\tag 3
$$

$$
\matrix
\delta_{\theta}M^A =\theta^A,\qquad
\delta_{\theta}B_{\mu\nu}=(-{i \over{2}}\ov{\theta}\Gamma_{\mu\nu}M
-{i\over{2}}\ov{M}\Gamma_{\mu\nu}\theta);
\endmatrix
\tag 4
$$
where $\Lambda$ is the local parameter associated to the gauge structure
group $U(1)$, $\epsilon$
and $ \theta$ are the infinitesimal parameters associated to
differentiable deformations in the space of $U(1)$ connections and of
sections of $S^+\otimes L$ respectively.

The field equations associated to (1) are
$$
F_{\mu\nu}+B_{\mu\nu}+ {i \over{2}} \ov{M} \Gamma_{\mu\nu} M = 0 ,\tag 5
$$

The gauge invariances of the action (1) allow the following partial gauge
fixing conditions
$$
B_{\mu\nu}^{+}\equiv{1\over{2}}(B_{\mu\nu} + {1\over{2}}
\epsilon_{\mu\nu\sigma\rho}B^{\sigma\rho}) = 0 ,  \tag 6
$$
and
$$
\align
&\Gamma^{\mu}\D_\mu M=0,\tag7
\endalign
$$
where $\Gamma^{\mu}\D_{\mu}$ is the Dirac operator that maps sections of
$S^{+}\otimes L$ to sections of $S^{-}\otimes L$.  It is here where we
need to dress up the 4-manifold X with a riemannian structure.
By using (6)
into (5) the field equations then reduce to

$$
\align
& F_{\mu\nu}^{-}+B_{\mu\nu}^{-}=0, \tag 8a \\
& F_{\mu\nu}^{+}+{i\over{2}}\ov{M}\Gamma_{\mu\nu}M=0 . \tag 8b
\endalign
$$
{}From (6) and (8a) we determine the auxiliarly field $B_{\mu\nu}$. The eqs.
(8b) and (7) are the monopole equations obtained in [1].

We are now going to construct the BRST invariant action following
standard procedures [5].
We consider the canonical analysis of our problem in one chart $ U_{L}$ of
the base manifold X. We shall show that the expression of the BRST charge
density allows a local treatment of this problem, ending up with a
covariant effective action which may be globally defined by patching
together the local expressions. This property is an important one since
we do not require any global decomposition of X into a product $R \times
\Sigma$ as is usually the case in an ordinary canonical formulation.

The canonical form of the action is
$$
\align
&S=\sum_{L} S_{L},
\endalign
$$
$$
\align
S_{L}=\int_{U_{L}}{}d^4x{}\;e_{L}[{}&\dot{A}_i\epsilon^{ijk}(F_{jk}+B_{jk}+
{i\over{2}}\ov{M}\Gamma_{jk}M)+A_0\D_i(\epsilon^{ijk}(F_{jk}+B_{jk}+
{i\over{2}}\ov{M}\Gamma_{jk}M))+ \\
&(B_{0i} + {i \over{2}}\ov{M}\Gamma_{0i}M)\epsilon^{ijk}
(F_{jk}+B_{jk}+ {i\over{2}}\ov{M}\Gamma_{jk}M)] , \tag 9
\endalign
$$
where $ e_{L}$ is the partition of unity. The eq. (9) yields the canonical
conjugate momenta to $A_i$ $$
\pi^i=\epsilon^{ijk}(F_{jk}+B_{jk}+{i
\over{2}}\ov{M}\Gamma_{ij}M).
$$

$A_0$ and $B_{0i}+ {i\over{2}}\ov{M}\Gamma_{0i}M $ are the  Lagrange
multipliers associated respectively to the constraints

$$
\align
&\phi \equiv \D_i\pi^i=0 ,\tag 10a\\
&\phi^i \equiv \pi^i=0 ,\tag 10b
\endalign
$$
the other constraints are given by
$$
\align
&\phi^A \equiv \eta_A =0,\\
&\ov{\phi}^A \equiv \ov{\eta}_A=0 ,\tag 10c
\endalign
$$
where $\eta_A$ and $\ov{\eta}_A$ are the conjugate momenta to $M^A$ and
$\ov{M}^A$ respectively.

All the constraints conmute, however (10a) and (10b) are not independent.
The reducibility matrix is given by
$$a\equiv{(\D_i,-1)}.\tag 11$$

To construct the BRST charge we follow Ref.[5] and introduce the minimal
sector of the extended phase space expanded by the conjugate pairs:

$$
(A_i,\pi^i);(M^A,\eta_A);(\ov{M}^A,\ov{\eta}_A);(C_1,\mu^1),
(C_{1i},\mu^{1i});(C_{11},\mu^{11});(C^A,\mu_A), \tag 12
$$
where we have introduced the ghost and antighost associated to the first
class constraints.

The off-shell nilpotent BRST charge is then given by:
$$
\align
&\Omega=\sum_{L}\Omega_{L}
\endalign
$$
$$
\Omega_{L}=\int_{U_{L}}d^4x\;e_{L}
(-(\D_iC_1)\pi^i+C_{1i}\pi^i-2iC^A \eta_A+2i\ov{C}^A \ov{\eta}_A
-(\D_iC_{11})\mu^{1i}-C_{11}\mu^1) ,\tag 13
$$
$\Omega_{L}$ acting on the configuration space satisfies
$\Omega^{2}_{L}=0$. This property may be directly checked from (18).

We now define the non minimal sector of the
extended phase space [5]. It contains extra ghosts, antighosts and Lagrange
multipliers. First we introduce the C-fields
$$
 C_m,C_{mi};\ \ \
C_{mn},C_{mni};\ \ \ \ \ \ m,n=1,2,3
$$
where at least one of the indices $m,n$ take the values 2 or  3. In
addition to these ghost, antighost and  Lagrange multiplier fields we
introduce the $\lambda$ and $\theta$ fields  (Lagrange multipliers),
also in the non minimal sector, $$
\align
&\lambda_1^0,\ \lambda_{1i}^0;\ \lambda_{1m}^0;\ \ m=1,2,3\\
&\lambda_{11}^1;\\
&\theta_1^0,\ \theta_{1i}^0;\ \theta_{1m}^0;\ \ m=1,2,3\\
&\theta_{11}^1.
\endalign
$$

In this notation the 1 subscripts denote ghost associated to a gauge
symmetry of the action, the 2 subscripts denote antighost associated to a
 gauge fixing  condition in the effective action and the 3 subscripts denote
Lagrange  multipliers associated to a gauge fixing condition.
The effective action is then given by:
$$
\align
S_{eff}=\int_{U_{L}} d^4x
[&\pi^i\dot{A}_i+\eta_A\dot{M}^A+\ov{\eta}_A\dot{\ov{M}}^A+\mu^1\dot{C}_{1}+
\mu^{1i}\dot{C}_{1i}+\\
&\mu^{11}\dot{C}_{11}+\mu_A\dot{C}^A+\ov{\mu}_A\dot{\ov{C}}^A+
\\&\wh{\delta}(\lambda_1^0\mu^1+\lambda_{1i}^0\mu^{1i}+
\lambda_{11}^1\mu^{11}+\lambda^A\mu_A+\ov{\lambda}^A\ov{\mu}_A)+L_{GF+FP}],
\tag14
\endalign
$$
where
$$
L_{GF+FP}=\wh{\delta}(C_2\chi_2+C_{2\mu\nu}\chi_2^{\mu\nu}
+C_2^{\dot{A}}\chi_{\dot{A}}+
\ov{C}_2^{\dot{A}}\ov{\chi}_{\dot{A}})+
\wh{\delta}(C_{12}\chi_{12})+
\wh{\delta}(\lambda_{12}^0\Lambda_2+
\theta_{12}^0\Theta_2),\tag15
$$
is the sum of the generalizations of the Fadeev-Popov and gauge fixing
terms. In Eq.(15) $\chi_2$, $\chi_2^{\mu\nu}$, $\chi_{\dot{A}}$ and
$\ov{\chi}_{\dot{A}}$ are the gauge fixing
functions associated  to the constraints (10), while $\chi_{12}$,
$\Lambda_2$ and $\Theta_2$ are gauge fixing functions associated to the
reducibility problem. They must
fix the longitudinal part of the fields $C_{1\mu}$,$\lambda_{1}^{0}$ and
$\theta_{1}^{0}$ . Notice that $C_{2\mu\nu}$ is self-dual. The BRST
transformation for the canonical variables is given by $$
\wh{\delta}Z=(-1)^{\epsilon_z}\{ Z,\Omega \},\tag 16
$$
where $\epsilon_z$ is the grassmanian parity of $Z$. The BRST
transformation of the variables of the non minimal sector are given in
Ref.[5]. After integration of
the auxiliarly sector we finally choose gauge fixing functions that may be
written in a covariant form as
 $$
\align
&\chi_2=\D_{\mu}A^\mu-{\alpha\over{2}}C_3,\\
&\chi_2^{\mu\nu}={1\over{2}}
\epsilon^{\mu\nu\sigma\rho}B_{\sigma\rho}+B^{\mu\nu},\\
&\chi_{12}=\D^{\mu}C_{1\mu}+{1\over{2}}
(-\ov{C}^A M_A + \ov{M}^AC_A),\\
&\chi_{\dot{A}}=
-{i\over{2}}\D_{A\dot{A}}\ov{M}^A+\ov{C}_{3\dot{A}},\\
&\ov{\chi}_{\dot{A}}=
-{i\over{2}}\D_{A\dot{A}}M^A+C_{3\dot{A}},
\tag 17 \endalign
$$
where $C_{1\mu}=(-\lambda_{11}^0,C_{1i})$.
After elimination of all conjugate momenta in the functional integral, the
BRST transformation rules of all the remaining geometrical objects are
covariant and take the form
$$\align
&\wh{\delta}A_{\mu}=-\D_{\mu}C_1+C_{1\mu},\\
&\wh{\delta}C_1=C_{11},\\
&\wh{\delta}C_{1\mu}=\D_{\mu}C_{11},\\
&\wh{\delta}C_{11}=0,\\
&\wh{\delta}C_2=C_3,\\
&\wh{\delta}C_3=0,\\
&\wh{\delta}C_{2\mu\nu}=C_{3\mu\nu},\\
&\wh{\delta}C_{3\mu\nu}=0,\\
&\wh{\delta}C_{12}=C_{13},\\
&\wh{\delta}C_{13}=0,\\
&\wh{\delta}C^A=0,\\
&\wh{\delta}M^A=-2iC^A,\\
&\wh{\delta}C_{2}^{\dot{A}}=C_{3}^{\dot{A}},\\
&\wh{\delta}C_{3}^{\dot{A}}=0,\\
\tag 18
\endalign
$$
$C_{2\mu\nu}$ and $C_{3\mu\nu}$ are self dual fields.
The BRST invariant action, once we have eliminated $B_{\mu\nu}$, may be
written as
$$
S=S_0+S_1+S_2+S_3 , \tag 19
$$
where
$$\align
&S_0 =\; {1\over{8}}<{1\over{2}}F^{+AB}F_{AB}^{+}+g^{\mu\nu}\D_\mu
\ov{M}^A \D_\nu M^A
+{1\over{4}}R\ov{M}^AM_A-{1\over{8}}
\ov{M}^{(A}M^{B)}\ov{M}_{(A}M_{B)}>,\tag 20 \\
&S_1=\; <-C_2^{\mu\nu}\D_{\mu}C_{1\nu}+C_{13}\D_{\mu}C_1^{\mu}
 +C_{12}\D_{\mu}\D^\mu C_{11}>,\tag 21\\
\endalign
$$
$$\align
S_2=\;
<&-{1\over{2}}C_2^{AB}(\ov{M}_{(A}C_{B)}+\ov{C}_{(A}M_{B)})
+\ov{C}_2^{\dot{A}}\D_{A\dot{A}}C^A-\ov{C}^A\D_{A\dot{A}}C_2^{\dot{A}}\\
&+{i\over{2}}(\ov{M}^AC_{1A\dot{A}}C_2^{\dot{A}}-\ov{C}_2^{\dot{A}}
C_{1A\dot{A}}M^A)
-{1\over{2}}C_{13}(\ov{C}^AM_A-\ov{M}^AC_A)\\&+2iC_{12}\ov{C}^AC_A
+{1\over{2}}\ov {M}^A\sigma_{A\dot{A}}^\mu(D_{\mu}C_{1})C_{2}^{\dot{A}}
-{1\over{2}}\ov{C}_{2}^{\dot{A}}\sigma_{A\dot{A}}^{\mu}(D_{\mu}C_1)M^A> ,
\tag 22 \endalign
$$
 and finally
$$
S_3=\; <C_3(\D_{\mu}A^{\mu}-{\alpha\over{2}}C_3)+C_2\D_{\mu}\D^\mu
 C_1-C_2\D_{\mu}C_1^{\mu} >. \tag 23 $$
where $<....>$ denotes integration on the 4-manifold X.
In these expressions we
have rewritten the objects with world indices in
terms of the corresponding ones with spinorial indices. We use the same
notation as in [1].
$S_0$ corresponds to the action used by Witten in deriving the vanishing
theorems in [1]. While $S_1+S_2+S_3$ are the contributions of the ghost
and antighost fields in order to have a BRST invariant action.
The action $S_0$ agrees with the bosonic sector of the gauge fixed action
proposed in
[6]. The difference in the explicit expression for the remaining
terms arises in
that the latter is invariant under  BRST transformations which
close modulo gauge transformations. While the action we
present is invariant under an off-shell nilpotent charge.
 In order to compare with the formulation in [2] and [6], one may perform
the change of variables:
$$
\align
&\psi_{\mu}=-iC_{1\mu},\\ &\phi=iC_{11},\\ &\eta=-C_{13},\\
&\lambda=-2iC_{12},\\ &\chi_{\mu\nu}=-C_{2\mu\nu},\\
&\mu^A=C^A, \\&v^{\dot{A}}=2iC_{2}^{\dot{A}}. \tag 24
\endalign $$

We show now how to obtain the SUSY algebra from our nilpotent BRST
algebra. Let us consider the transformation law for $A_{\mu}$. We define
the SUSY transformation by
$$
\delta A_{\mu} :=\wh{\delta}A_{\mu}\;\mid_{C_{1}=0}
$$
we then have from (18)
$$
\align
\delta A_{\mu}&=C_{1\mu},\\
\delta \delta A_{\mu}&=\delta C_{1\mu}=\D_{\mu}C_{11}.\\
\endalign
$$

The SUSY algebra thus closes up to a gauge transformation generated by
$C_{11}$ as required.

The SUSY transformation for $M^A$ may be obtained by considering an
equivalent BRST formulation to (13). Instead of considering the
constraint (10a) we may take equivalently
$$
\D_{i}\pi^i+M^{A}\eta_{A}+\ov{M}^{A}\ov{\eta}_{A}=0
$$
The associated BRST charge is then given by
$$
\align
\Omega_{L}=\int_{U_{L}}d^4x\;e_{L}
(&-(\D_iC_1)\pi^i+C_{1i}\pi^i-2iC^A \eta_A+2i\ov{C}^A \ov{\eta}_A
-(\D_iC_{11})\mu^{1i}-C_{11}\mu^1 \\&+C_{1}M^A \eta_A+C_{1}\ov{M}^A
\ov{\eta}_A-{i\over{2}}C_{11}M^A \mu_A \\&+{i\over{2}}C_{11}\ov{M}^A
\ov{\mu}_A+C^{A}\mu_{A}C_{1}+\ov{C}^{A}\ov{\mu}_{A}C_{1})
\endalign
$$
comparing with (13) we see some other terms coming from the new choice of
constraints. The BRST charge is again nilpotent when acting on the
configuration space of the fields after the elimination of the auxiliary
ones. The nilpotent BRST transformation laws are now
$$
\align
\wh{\delta}M^A&=-2iC^A+C_{1}M^{A},\\
\wh{\delta}C^A&=-{i\over{2}}C_{11}M^{A}+C^{A}C_{1},
\endalign
$$
there are analogous changes for $\ov{M}^A$ and $\ov{C}^A$, while the
transformation law for the other fields are as in (18).
We define as before the SUSY transformations of $M^A$ and $C^A$. We have
$$
\align
\delta M^{A} &:=\wh{\delta}M^{A}\;\mid_{C_{1}=0}\\
\delta C^{A} &:=\wh{\delta}C^{A}\;\mid_{C_{1}=0}
\endalign
$$
We then obtain
$$
\align
\delta\delta M^{A} &=-2i\delta C^{A}=-C_{11}M^{A}\\
\delta\delta C^{A} &=-{i\over{2}}C_{11}\delta M^{A}=-C_{11}C^{A}
\endalign
$$
as required.
As shown the full SUSY algebra results from our nilpotent BRST charge.
The combination of constraints we have considered corresponds to a
canonical change of coordinates in the original symplectic geometry.

In summary, we introduced a topological action with a large class of
local symmetries, whose field equations are the Seiberg-Witten monopole
equations found in [1]. By following a covariant gauge fixing procedure
we obtained a covariant BRST invariant effective action . The BRST
generator obtained is nilpotent off-shell. The canonical construction of
the nilpotent BRST charge has been carried out without any further
requirements on the base manifold beyond those assumed for the set up of
action (1). This construction uses particular properties of the BRST
charge for this topological theory. Finally we show
how the twisted  N=2 supersymmetric algebra used to get the
Seiberg-Witten topological theory may be directly obtained from our
nilpotent BRST charge. This last result shows that the supersymmetry is
hidden within the BRST symmetry and it seems  not to be  the main
ingredient in the whole Seiberg-Witten construction.

\vskip 1cm
Acknowledgements: we would like to thank L. Recht for very helpful and
interesting discussions. We also would like to thank P. Fre and R.
Dijkgraaf for conversations. Finally we thank ICTP centre for hospitality
during the Spring School when this work was done.

\vskip 0.5cm
\noi
{\bf REFERENCES}
\vskip .3cm

\item{[1]}N. Seiberg and E. Witten, {\it Nucl. Phys.} {\bf B426}
(1994) 19.
\item{}N. Seiberg and E. Witten, hep-th/9408099.
\item{}E. Witten, hep-th/9411102.
\item{[2]}E. Witten, {\it Commun. Math. Phys.} {\bf 117}
(1988) 353.
\item{[3]}R. Gianvittorio, A. Restuccia and J. Stephany, {\it Phys.
Lett.} {\bf B347} (1995) 279.
\item{}J.M.F. Labastida and M. Pernici, {\it Phys. Lett.} {\bf B212}
(1988) 56.
\item{}L. Baulieu and I. M. Singer, {\it Nucl. Phys.} (Proc.
Suppl.) {\bf 5B} (1988) 12.
\item{}Y. Igarashi, H. Imai, S. Kitakado and H. So, {\it Phys. Lett.}
{\bf B227} (1989) 239.
\item{}C. Arag\~{a}o and L. Baulieu, {\it Phys. Lett.}
{\bf B275} (1992) 315.
\item{[4]}F. Hirzebruch and H. Hopf, {\it Math. Ann.} {\bf 136} (1958) 156.
\item{[5]}M. I. Caicedo and A. Restuccia, {\it Class. Quan. Grav} {\bf 10 }
(1993) 833;
\item{}M. I. Caicedo and A. Restuccia, {\it Phys. Lett.} {\bf B307 } (1993) 77.
\item{[6]}J.M.F. Labastida and Mari\~no, SLAC-PUB-US-FT 3/95.

\end